# Unusual enhancement of effective magnetic anisotropy with decreasing particle size in maghemite nanoparticles

K. L. Pisane, Sobhit Singh and M. S. Seehra*

Department of Physics & Astronomy, West Virginia University, Morgantown, WV 26506, USA.

-----------------------------------------------------------------------------------------------------------------

## Abstract

Experimental results and a model are presented to explain the observed unusual enhancement of the effective magnetic anisotropy $K_{eff}$ with decreasing particle size D from 15 nm to 2.5 nm in $\gamma$-$Fe_2O_3$ nanoparticles (NPs). The samples include oleic acid-coated NPs with D = 2.5, 3.4, 6.3 and 7.0 nm investigated here, with the results on other sizes taken from literature. $K_{eff}$ is determined from the analysis of the frequency dependence of the blocking temperature $T_B$ after considering the effects of interparticle interactions on $T_B$. The data of $K_{eff}$ vs. D is fit to the derived core-shell-based relation: $K_{eff} = K_b + (6K_S/D) + K_{sh}\{[1-(2d/D)]^{-3} -1\}$, with $K_b$ = 1.9 x$10^5$ ergs/$cm^3$ as the bulk-like contribution of the core, $K_S$ = 0.035 ergs/$cm^2$ as the contribution of the surface layer, and $K_{sh}$ = 1.057 x$10^4$ ergs/$cm^3$ as the contributions of spins in the shell of thickness $d$ = 1.1 nm. This equation represents an extension of the often-used Eq.: $K_{eff} = K_b + (6K_S/D)$. Significance of this new result is discussed.

*Corresponding author. Email: mseehra@wvu.edu; Phone: 304-293-5098

There continues to be world-wide interest in the size dependent magnetic properties of magnetic nanoparticles (NPs) and their various applications in diverse fields such as catalysis, ferrofluids, sensors, magnetic storage media, and biomedicine [1-6]. Based on the nature of magnetic exchange coupling, two classes of NPs may be distinguished: magnetic NPs of metals such as Co [7,8] and Fe [9] and those of oxides such as antiferromagnetic CuO [10] and NiO [11] and ferrimagnetic magnetite ($Fe_3O_4$) [12,13] and maghemite ($\gamma$-$Fe_2O_3$) [14-17]. In the oxides, the exchange coupling between the transition-metal ions is facilitated by super-exchange via the intermediate $O^{2-}$ ions. It is now generally recognized that magnetic properties of NPs depend on several factors such as particle size D (or volume V), size distribution, morphology, interparticle interactions (IPI), and interaction between surface spins and ligands. To unravel all these different effects even in structurally well-characterized samples requires detailed measurements as a function of particle size, temperature, magnetic field strength and measurement frequency.

An important parameter for magnetic NPs is their magnetic anisotropy energy $E_a = K_{eff} V$ which keeps the particle magnetic moment aligned in a particular direction. Therefore, how this energy and the anisotropy constant $K_{eff}$ varies with particle size D is important in relation to the stability of the stored information, e.g. in recording media. In addition to the finite size-effects, spins on the surface experience different anisotropy because of the reduced dimensionality and broken exchange bonds. The latter is particularly valid for oxides where presence of oxygen vacancies on the surfaces disrupts the super-exchange coupling between the cations [18]. In this paper, we present detailed magnetic measurements in structurally well-characterized $\gamma$-$Fe_2O_3$ NPs of diameter D = 2.5, 3.4, 6.3 and 7.0 nm coated with oleic-acid (OA), with the analysis of the results focusing on the D-dependence of the effective anisotropy $K_{eff}$ and energy barriers to magnetic moment rotation. By including similar data available from literature on other sizes of $\gamma$-$Fe_2O_3$ NPs with D up to 15 nm, it is shown that an extension of the often used Eq.: $K_{eff} = K_b + (6K_S/D)$ [9, 19] to $K_{eff} = K_b + (6 K_S/D) + K_{sh} \{[1-(2d/D)]^{-3} -1\}$ is needed to explain the variation of $K_{eff}$ vs. D, particularly for smaller D with $K_b$ ($K_S$) as the bulk (surface) anisotropy constants and $K_{sh}$ as the anisotropy of spins in the shell of thickness $d$.

The theoretical basis for determining the effective anisotropy $K_{eff}$ from experimental data is first outlined here. For non-interacting single-domain magnetic NPs, each with volume V, $K_{eff}$ is related to the experimentally observed blocking temperature $T_B$ by the relation:

$$T_B = (K_{eff}V) / [k_B \ln(f_o/f_m)] \text{ ---------------------- (1)}$$

Here $f_o \sim 10^{10}$ Hz is the system-dependent attempt frequency [20, 21], $f_m$ is the measurement frequency and $k_B$ is the Boltzmann constant. In the presence of IPI characterized by an effective temperature $T_o < T_B$, Eq. (1) is replaced by [22, 23]:

$$T_B = T_o + (K_{eff}V) / [k_B \ln(f_o/f_m)] \text{ --------------- (2)}$$

To determine $T_o$, $T_B$ is measured at several $f_m$, analysis of which can determine $T_o$, $K_{eff}$ and the attempt frequency $f_o$ [24]. The presence of IPI can also be gauged by $\Phi = \Delta T_B / [T_B \Delta \log_{10} f_m]$ with $\Delta T_B = T_B(2) - T_B(1)$ being the difference in $T_B$ determined at two sufficiently different measuring frequencies $f_m(2) > f_m(1)$ [25]. $\Phi < 0.13$ signifies presence of IPI with their strength increasing with decreasing $\Phi$. $\Phi$ and $T_o$ are related by the Eq. [26]:

$$\Phi = \Phi_o[1 - (T_o/T_B(1))] \text{ -------------- (3)}$$

In Eq. (3), $\Phi_o = 2.3026/\{\ln[f_o/f_m(2)]\}$. As noted above, measurements of $T_B$ at several $f_m$ is needed to determine $K_{eff}$ [24]. For $\gamma$-Fe$_2$O$_3$ NPs, such data is reported for few sizes [17, 27-32] which is used here along with our new results on the $\gamma$-Fe$_2$O$_3$ NPs with D=2.5, 3.4, 6.3 and 7.0 nm to discuss the size dependence of $K_{eff}$ for $\gamma$-Fe$_2$O$_3$ NPs.

An important parameter in NPs is the surface to volume ratio which, for spherical particles with a single layer of surface atoms, is 6/D. Since atoms on the surface of a NP experience different anisotropy $K_S$, the total $K_{eff}$ has been suggested to be given by [9]:

$$K_{eff} = K_b + (6K_S)/D \text{ ------------- (4)}$$

Eq. (4) was found to be valid for Fe NPs [9] for D > 4 nm. A core-shell model was used by Millan et al. [15] for $\gamma$-Fe$_2$O$_3$ NPs and Dutta et al. [12] for Fe$_3$O$_4$ NPs to explain the particle size dependence of the saturation magnetization $M_S$ following the easily derived equation: $M_S = M_S(b)[1-(2d/D)]^3$ where $d$ is the thickness of the shell containing disordered spins not contributing to $M_S$, and $M_S(b)$ is the saturation magnetization for bulk sample representing the core-spin contribution. Recent Monte-Carlo simulations applied to $\gamma$-Fe$_2$O$_3$ NPs have shown that the spin disorder starting at the surface gradually propagates inwards with decrease in D [33]. This effect creates a shell of thickness $d$ in which ordering of the spins is different from that of the surface spins and bulk-like ordering of the core spins. So we describe the particle using a core-shell geometry and propose extension of Eq. (4) to include the effect of shell of thickness $d$ yielding:

$$K_{eff} = K_b + (6 K_S/D) + K_{sh} \{[1-(2d/D)]^{-3} -1\} \text{------------------ (5)}$$

The validity of Eq. (5) is limited to $D > 2d$ since only in this limit the NPs have a core. The factor $\{[1-(2d/D)]^{-3} -1\}$ in the last term of Eq. (5) represents the ratio of shell-volume to core volume and it represents the fraction of the spins in the shell with effective anisotropy $K_{sh}$ different from $K_S$ and $K_b$. The validity of Eqs. (4) and (5) is tested later for $\gamma$-$Fe_2O_3$ NPs with D = 2.5 nm to 15 nm.

Nanoparticles of OA-coated $\gamma$-$Fe_2O_3$ were synthesized using a modification of the procedure of Hyeon et al [34] with details given in [35] and in the supplemental information (SI) [36]. The size distributions of the NPs determined from TEM micrographs are shown in Fig. 1 [36]. The average sizes determined from this analysis are: D = 7.0 ± 0.8, 6.3 ± 0.6, 3.4 ± 0.8 and 2.5 ± 0.7 nm. The TEM images (Fig.1) show round cross-sections of all NPs indicating nearly spherical NPs. From the x-ray diffraction patterns shown in Fig. 2 [36], the crystallite size D and strain $\eta$ were determined by fitting the instrument-corrected width $\beta$ of the lines to the Williamson – Hall Eq.: $\beta \cos\theta = (0.9\lambda/D) + \eta \sin\theta$. The linear plots of $\beta \cos\theta$ vs. $\sin\theta$ given in [35] yielded D = 6.7, 6.3, 4.1, and 2.9 nm and $\eta(10^{-3})$ = 5, 0.7, 15, and 4 in agreement with D = 7.0, 6.3, 3.4, and 2.5 nm determined from TEM.

Magnetic properties of the four $\gamma$-$Fe_2O_3$ NPs were measured using a Physical Properties Measurement System (Quantum Design, Inc). The temperature dependence of the ac magnetic susceptibilities, $\chi'$ and $\chi''$, for the four samples were measured at nine $f_m$ between 10 Hz and 5 kHz with applied $H_{ac}$ = 5 Oe and focusing on the region around $T_B$, with the representative data for D = 3.4 nm $\gamma$-$Fe_2O_3$ given in Fig. 3. Similar data for other sizes are available in [35]. The data show a peak in $\chi''$ which defines $T_B$ and it shifts to higher T with increase in $f_m$. The temperature dependence of $\chi'$ is such that the relation $\chi'' \sim d(\chi'T)/dT$ is valid [24].

The results of the change in $T_B$ with change in $f_m$ (see Fig. S1) for the four samples were analyzed using Eq. (2) with $f_o$ = 2.6 × $10^{10}$ Hz determined for the 7.0 nm sample which showed negligible IPI since $T_o$ = 0 K was found to be valid for this case [17] and confirmed by the measured value of $\Phi$ = 0.12 for this sample. For the other three samples with D = 6.3, 3.4 and 2.5 nm, $\Phi$ = 0.084, 0.094, and 0.080 respectively is determined implying the presence of significant IPI and so $T_o > 0$. For these three samples, assuming $f_o = 2.6 \times 10^{10}$ Hz to be the same, $T_o$ was then varied so that the plots of $(T_B-T_o)^{-1}$ versus $lnf_m$ yielded straight lines as shown in Fig. 4 according to Eq. (2). The negative slope of the straight line equals ($k_B/K_{eff}V$) and the intercept equals $k_B lnf_o/K_{eff}V$. The solid lines are fits to Eq. (2) with $T_o$ = 0, 11, 2.5, and 12.5 K and $K_{eff}$ (in units of $10^5$ ergs/cm$^3$) = 5.6 ± 1.8, 7.5 ± 1.0, 18.6 ± 10.0, and 80 ± 28 for the 7.0, 6.3, 3.4 and 2.5 nm NPs, respectively. For volume V, particles were assumed to be spheres of diameter D (see Fig. 1).

It is noted that at 10 Hz, the measured $T_B$ = 35, 42, 21 and 29 K for the D = 7.0, 6.3, 3.4, and 2.5 nm NPs respectively does not vary systematically with decrease in D because of the different magnitude of $T_o$. However, using the magnitudes of $T_o$ given above it is easily seen that there is a systematic change in ($T_B-T_o$) with size D as expected, signifying the importance of accurately determining $T_o$ as shown here (see Fig. S2).

The variation of $K_{eff}$ determined above with particle size D for the four NPs of $\gamma$-Fe$_2$O$_3$ is plotted in Fig. 5 along with suitable data available from the literature on different sized $\gamma$-Fe$_2$O$_3$ NPs [27-32]. For example, Laha et al. [31] investigated the properties of 6.0 nm $\gamma$-Fe$_2$O$_3$ NPs providing $T_o$ = 50 K and ($K_{eff}V/k_B$) = 350 K, the latter yielding $K_{eff}$ = 4.3 x $10^5$ ergs/cm$^3$. For $\gamma$-Fe$_2$O$_3$ NPs with D = 4.8, 6.5, 7.7, 9.3 and 10.0 nm, Gazeau et al. [28] reported magnitudes of volume (surface) anisotropy $K_V$ ($K_S$) which we used to determine $K_{eff}$ following Eq. (4) since these particles have D > 4nm (see discussion later). For the D = 4.0 nm $\gamma$-Fe$_2$O$_3$ NPs, Nadeem et al. [29] reported $T_o$ = 46 K and ($K_{eff}V/k_B$) = 203 K, the latter yielding $K_{eff}$ = 8.36 x $10^5$ ergs/cm$^3$. The data provided by Komorida et al. [32] for the D = 5.1 nm $\gamma$-Fe$_2$O$_3$ NPs was used to determine $K_{eff}$ = 7.0x $10^5$ ergs/cm$^3$. Finally, the magnitudes of $K_{eff}$ reported by Fiorani et al. [27] for the D = 2.7, 4.6 and 8.7 nm $\gamma$-Fe$_2$O$_3$ NPs and by Demchenko et al. [30] for the D = 10.5, 13.3 and 15.3 nm $\gamma$-Fe$_2$O$_3$ NPs were used in making the plot of $K_{eff}$ vs. D in Fig. 5.

First to check Eq. (4), the data are plotted as a variation of $K_{eff}$ versus 1/D in Fig. 6 where it is evident that for the NPs with D < 5 nm, there are strong deviations from the linearity of $K_{eff}$ vs. 1/D expected from Eq. (4). However, for D > 5 nm, the data follows the linearity predicted by Eq. (4) quite well with the slope yielding $K_S$ = 0.035 ergs/cm$^2$ and the intercept $K_b$ = 1.9 x $10^5$

ergs/cm$^3$. Next the same data are fit to Eq. (5) with the results also shown in Fig. 6 using $K_S$ = 0.035 ergs/cm$^2$ and $K_b$ = 1.9 x10$^5$ ergs/cm$^3$ determined above and using $d$ and $K_{sh}$ as fitting parameters. The best fit is obtained with shell thickness $d$ = 1.1 nm and $K_{sh}$ = 1.057 x10$^4$ ergs/cm$^3$ and this fit based on Eq. (5) with the parameters listed above is also shown in Fig. 6 and in Fig. 5. This magnitude of $d$ = 1.1 nm agrees quite well with $d$ ~ 1.0 nm determined by Millan et al. [15] from the particle size dependence of the saturation magnetization $M_S$ of γ-Fe$_2$O$_3$ NPs.

Comparison of Eq. (4) and (5) and their fits to the data shows that Eq. (4) is a special case of Eq. (5) for $d$ << D since for size D > 5 nm in Fig. 6, Eq. (4) provides a good description of the variation. However, for D < 5 nm, deviations from Eq. (4) becoming increasingly significant with decreasing D and the proposed Eq. (5) is needed to describe the observed variation of $K_{eff}$ vs. size D. Based on the core-shell model with the above listed magnitudes of $K_b$, $K_S$, $K_{sh}$ and $d$ = 1.1 nm, it is easily shown that at D = 3 nm, the contribution of the $K_{sh}$ term of Eq. (5) to $K_{eff}$ is about 38%, decreasing to 13% for D = 4 nm, to 3.7% for D = 8 nm and to 2% for D = 15 nm. However, the contribution of the $K_S$ term remains significant even for D = 20 nm. Therefore, the $K_{sh}$ term becomes important only for sizes D < 5 nm and for such cases, Eq. (5) proposed here needs to be used. For morphologies different from a sphere, the factor 6 in Eqs. (4) and (5) should be replaced by a factor representing that particular morphology. The validity of Eq. (5) can be tested in other NP systems where similar data of $K_{eff}$ vs. D becomes available.

**Acknowledgement:** This work was supported in part by a grant from the U.S. National Science Foundation, Grant # DGE-1144676.

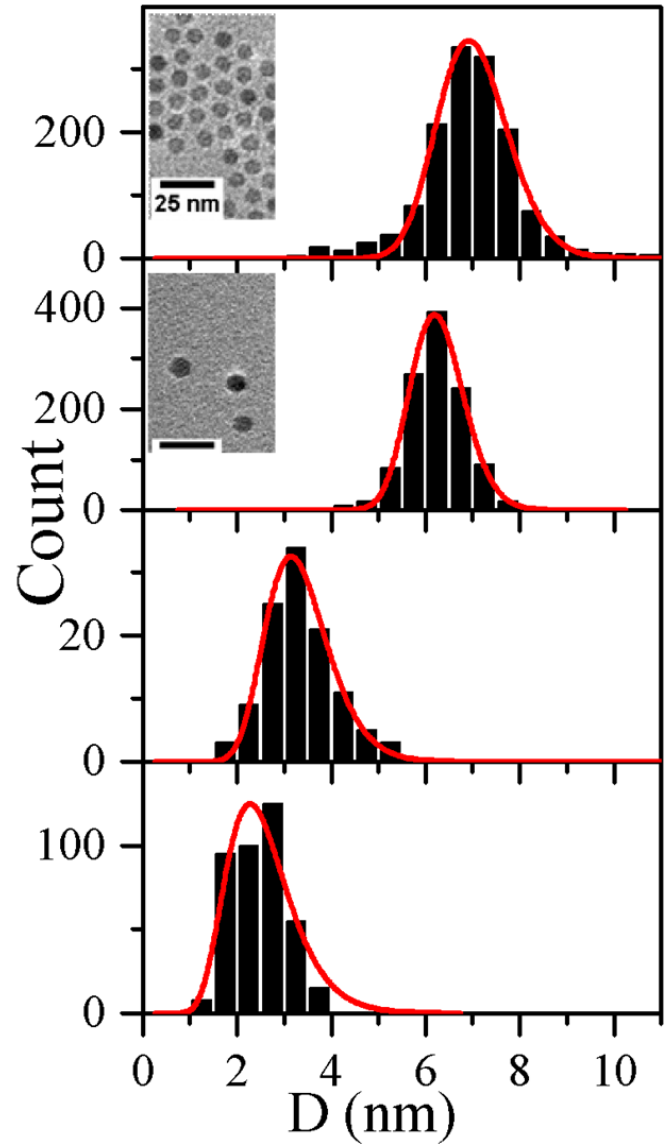

Fig. 1. Size distributions determined from TEM micrographs for the 7.0 nm, 6.3 nm, 3.4 nm and 2.5 nm γ-Fe$_2$O$_3$ NPs (top to bottom). The vertical black bars represent the number of particles counted with the corresponding diameter and the red curves are fits to log-normal distributions. The insets show representative TEM micrographs for each particle size.

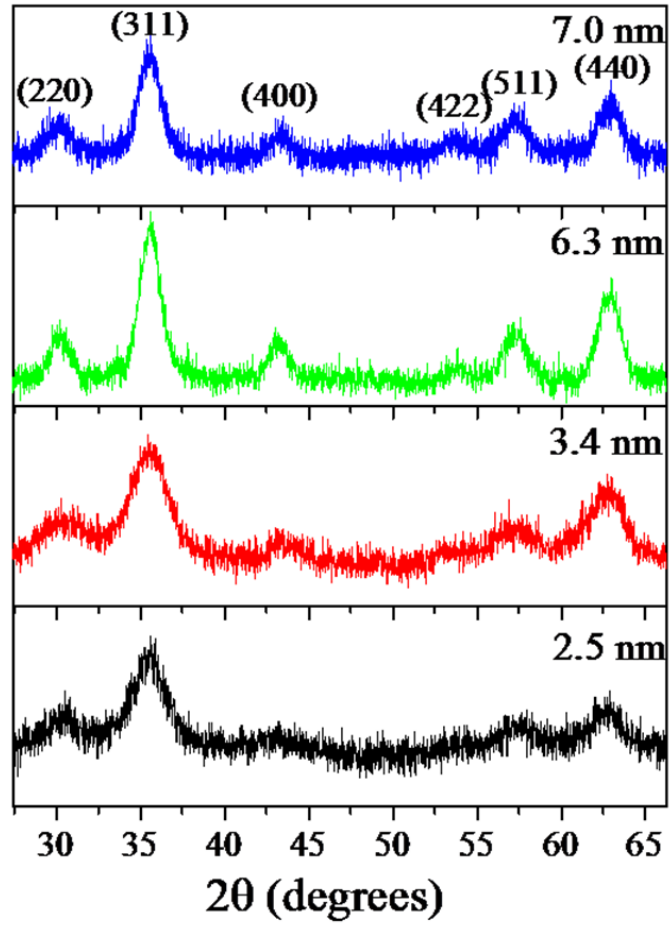

Fig. 2. XRD patterns of the four γ-$Fe_2O_3$ NPs with the Miller indices of the lines listed above the peaks in the top panel.

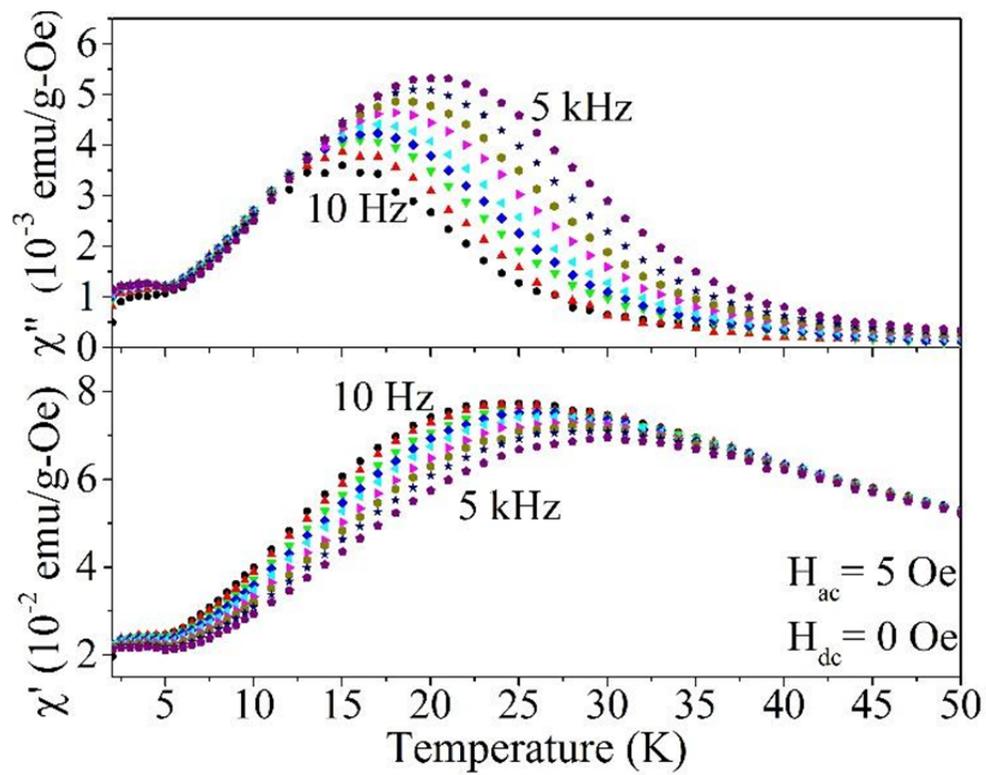

Fig. 3. Temperature dependence of the ac magnetic susceptibilities of the 3.4 nm γ-$Fe_2O_3$ NPs for frequencies $f_m$ =10 Hz, 20 Hz, 50 Hz, 100 Hz, 200 Hz, 500 Hz, 1 kHz, 2 kHz, and 5kHz.

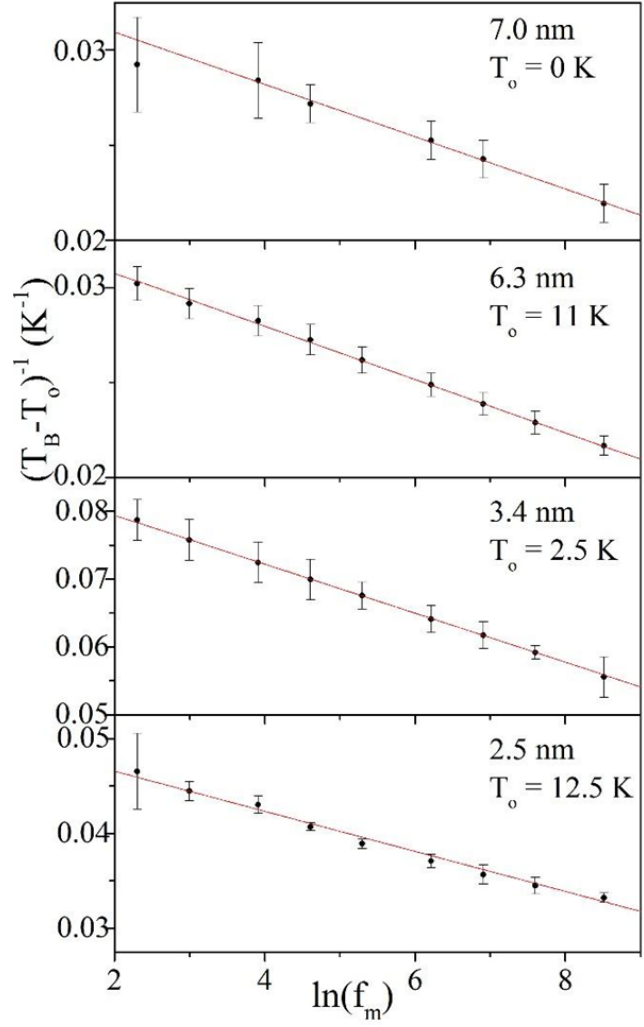

Fig. 4 Plots of $(T_B - T_o)^{-1}$ vs. $ln(f_m)$ with the solid red lines as fits to Eq. (2) using parameters determined from the fits given in the text.

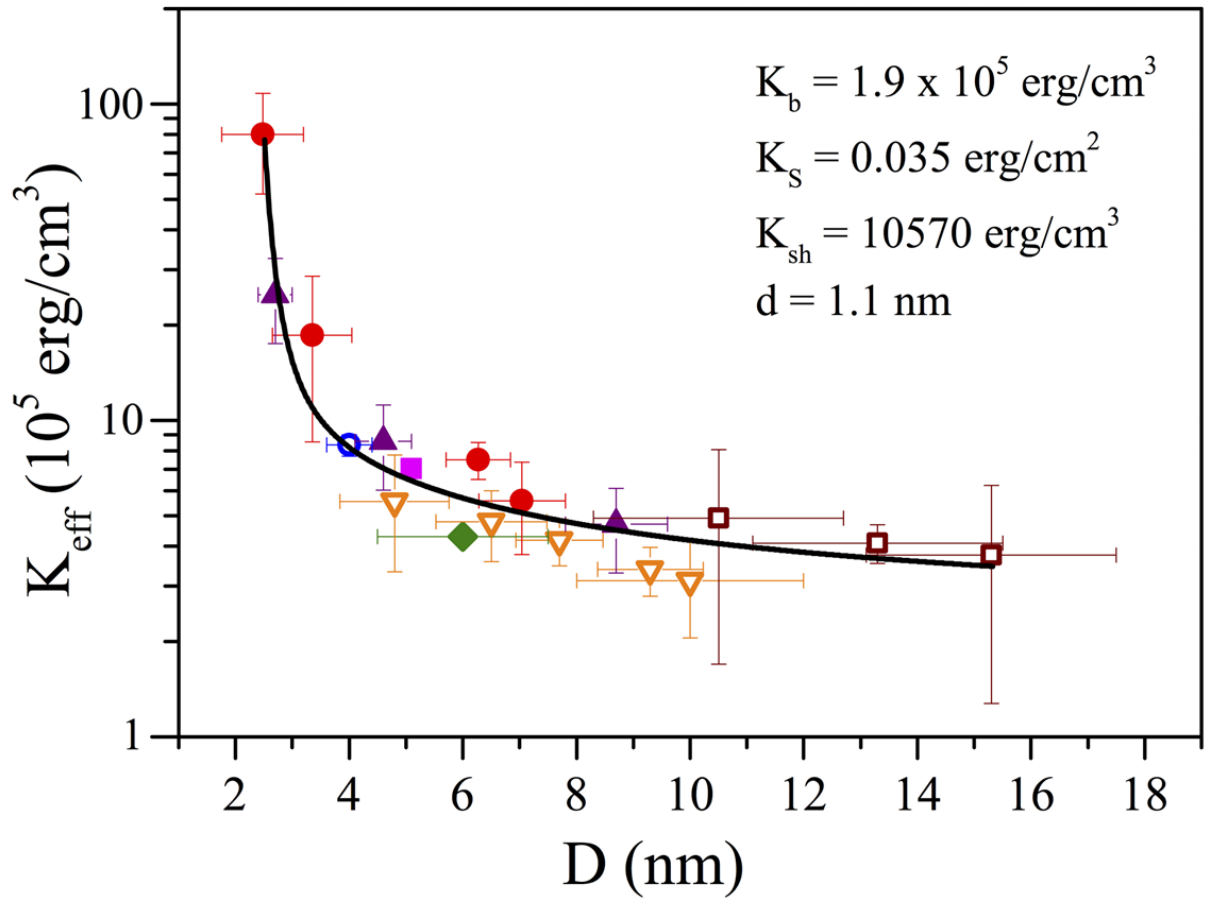

Fig. 5. Variation of the effective anisotropy constant $K_{eff}$ (in log-scale to highlight the data for larger particles) with particle size D. Different symbols represent data from different groups as follows: solid purple triangles (Fiorani, et al., 2002), open blue circle (Nadeem, et al., 2011), solid green diamond (Laha, et al., 2014), open orange triangles (Gazeau, et al., 1998), solid pink square (Komorida, et al., 2009), open brown squares (Demchenko, et al., 2015); solid red circles (this work). The solid line is fit to Eq. (5) with the parameters listed in the figure.

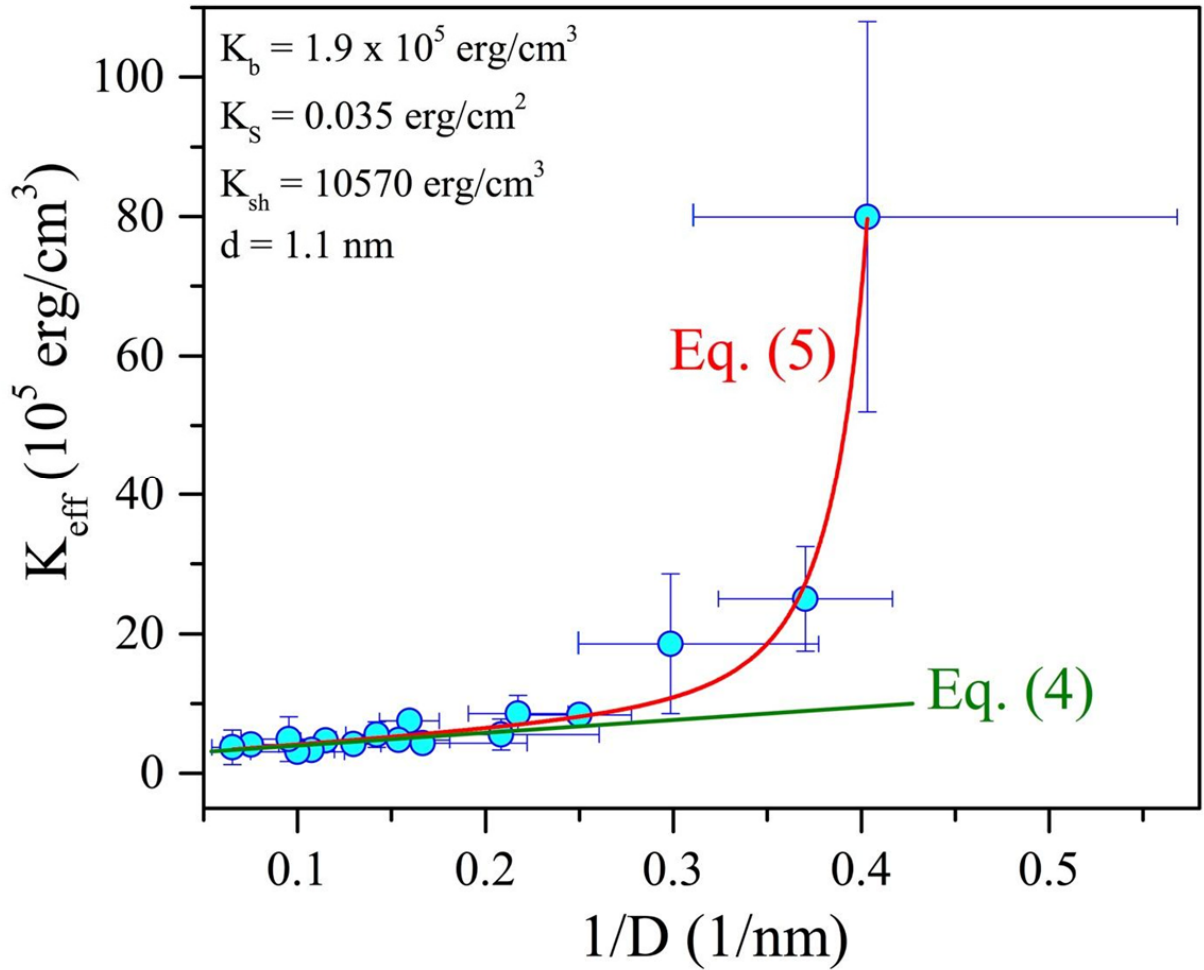

Fig. 6. The data of Fig.5 is replotted here as $K_{eff}$ versus $1/D$. The fits to Eq. (4) and Eq. (5) are shown as solid lines using the parameters listed in the figure.

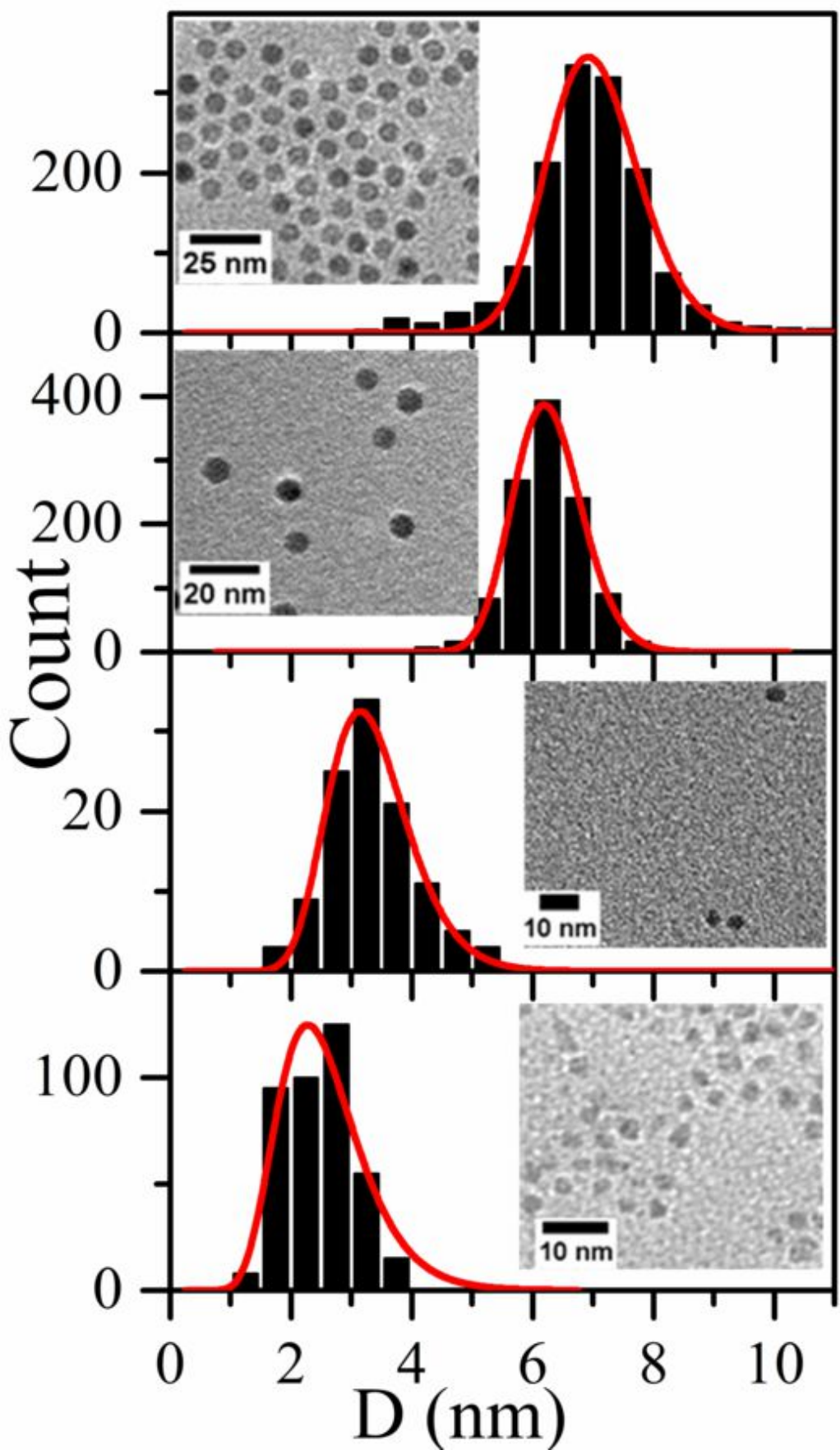

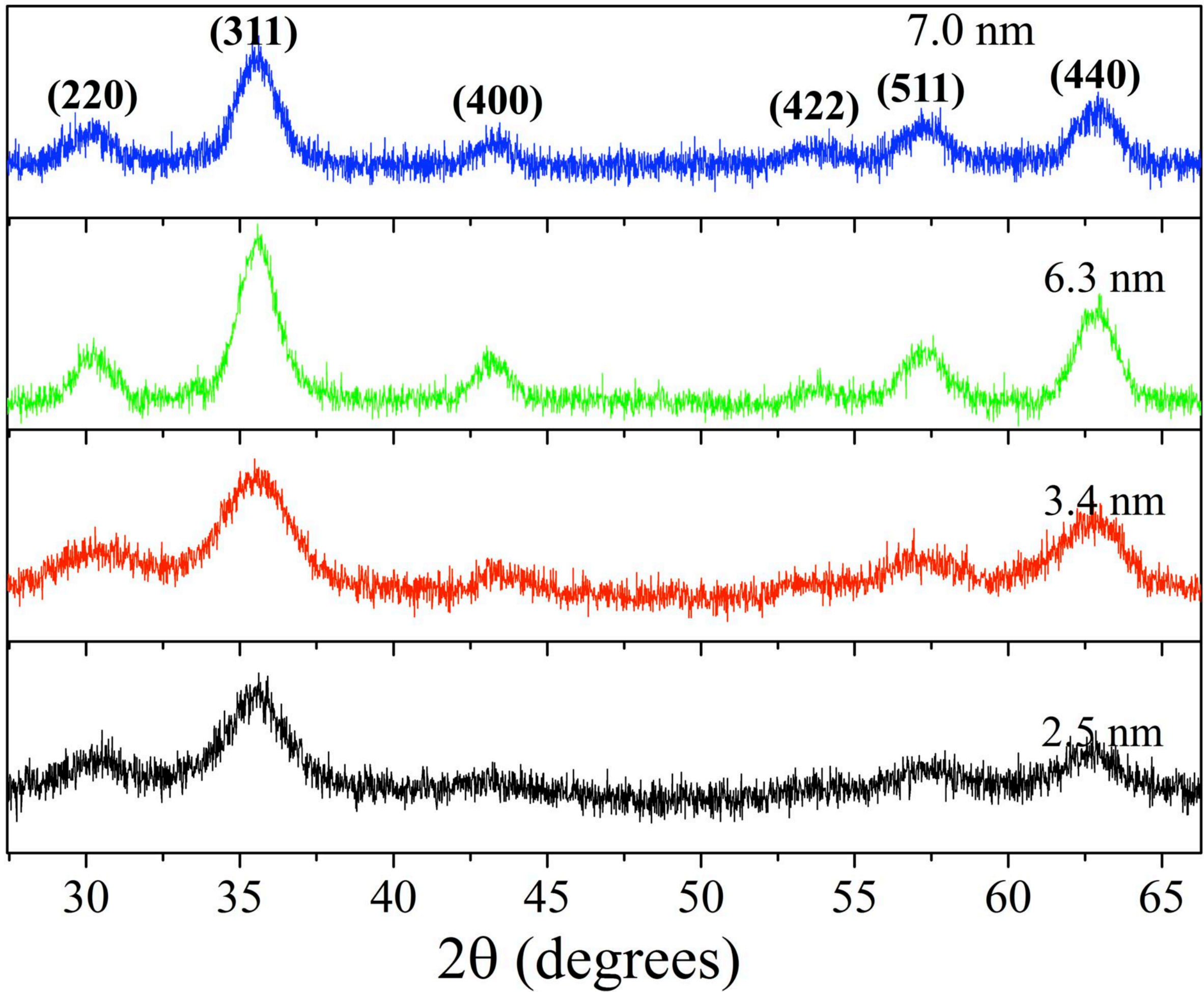

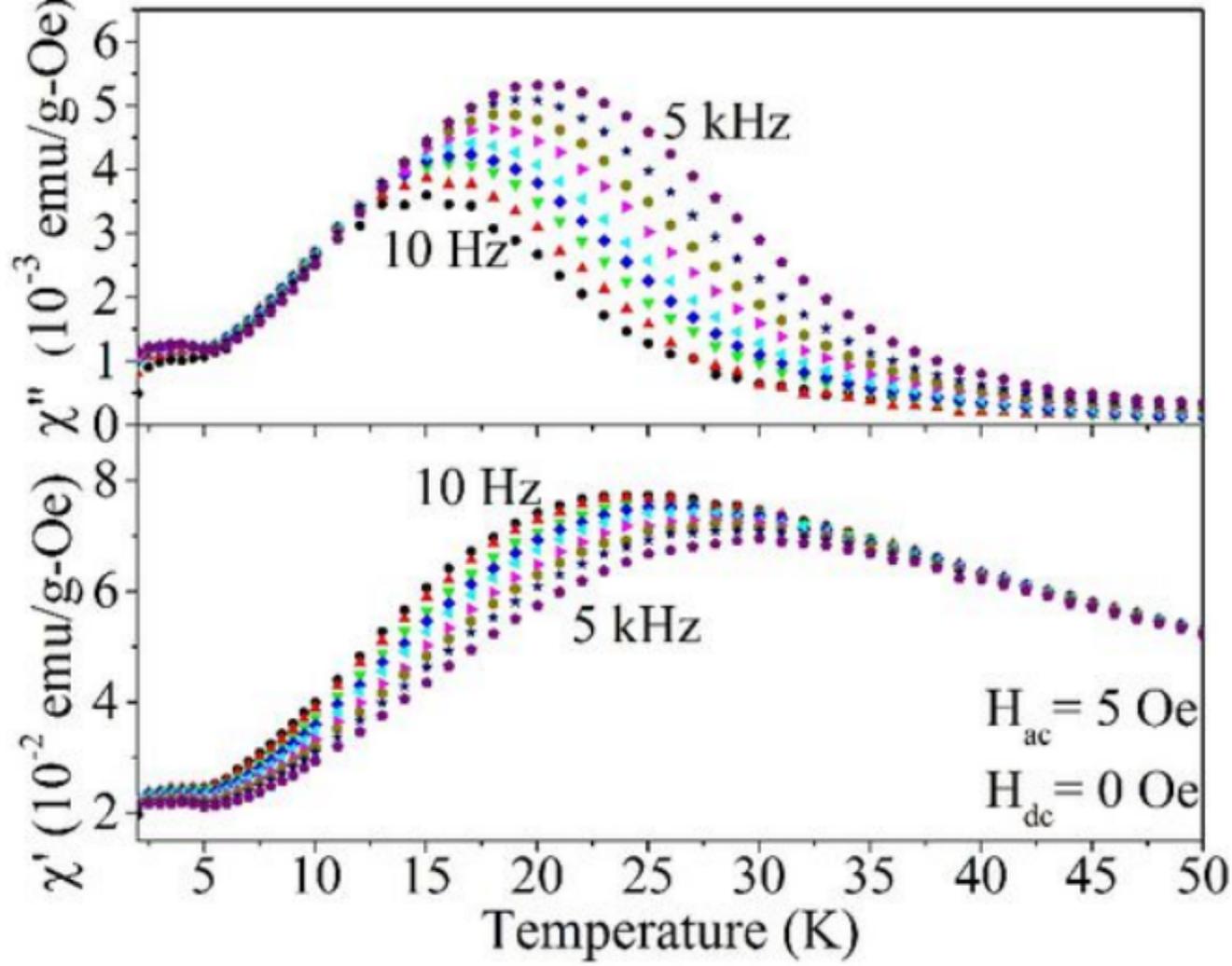

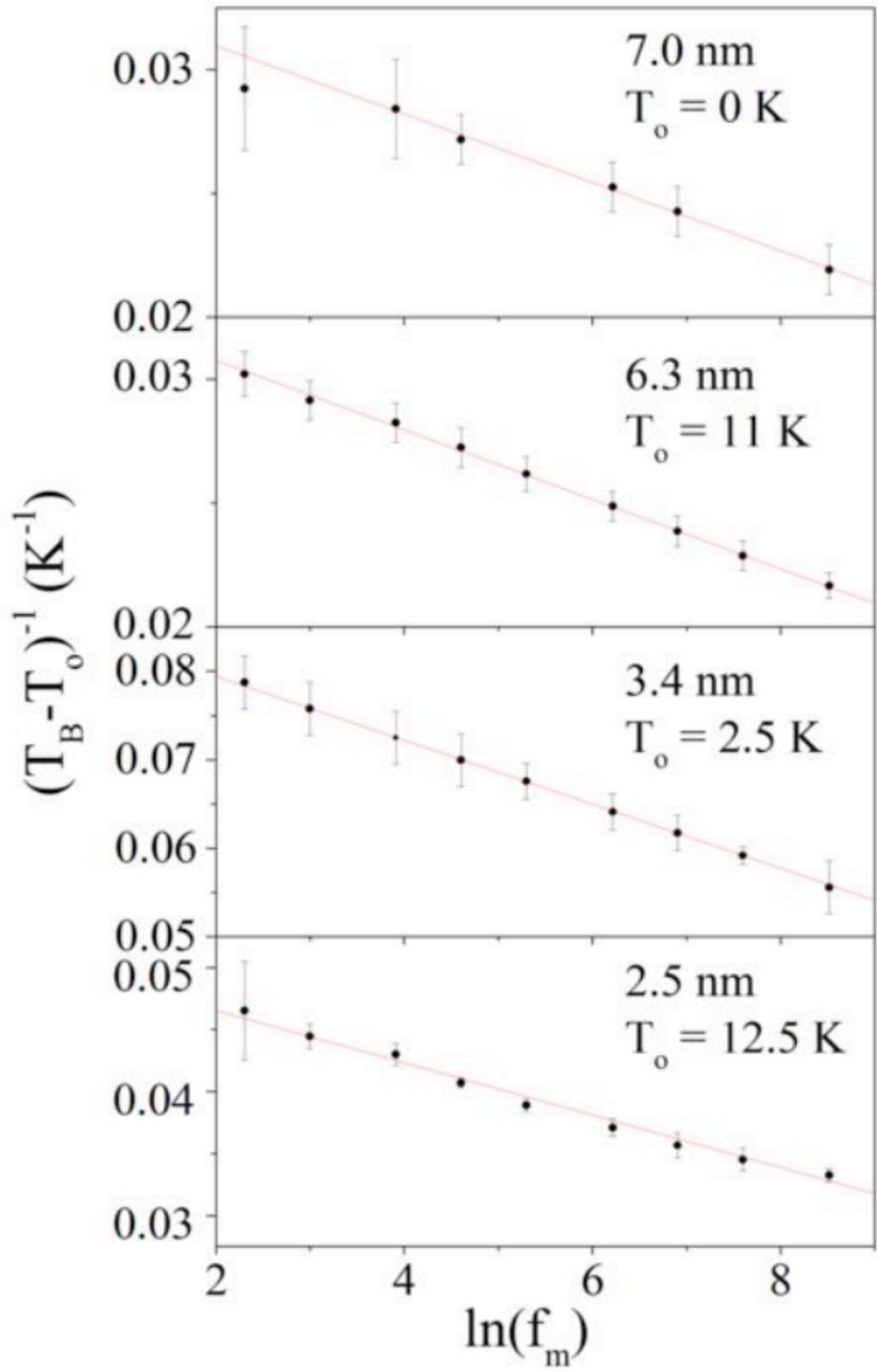

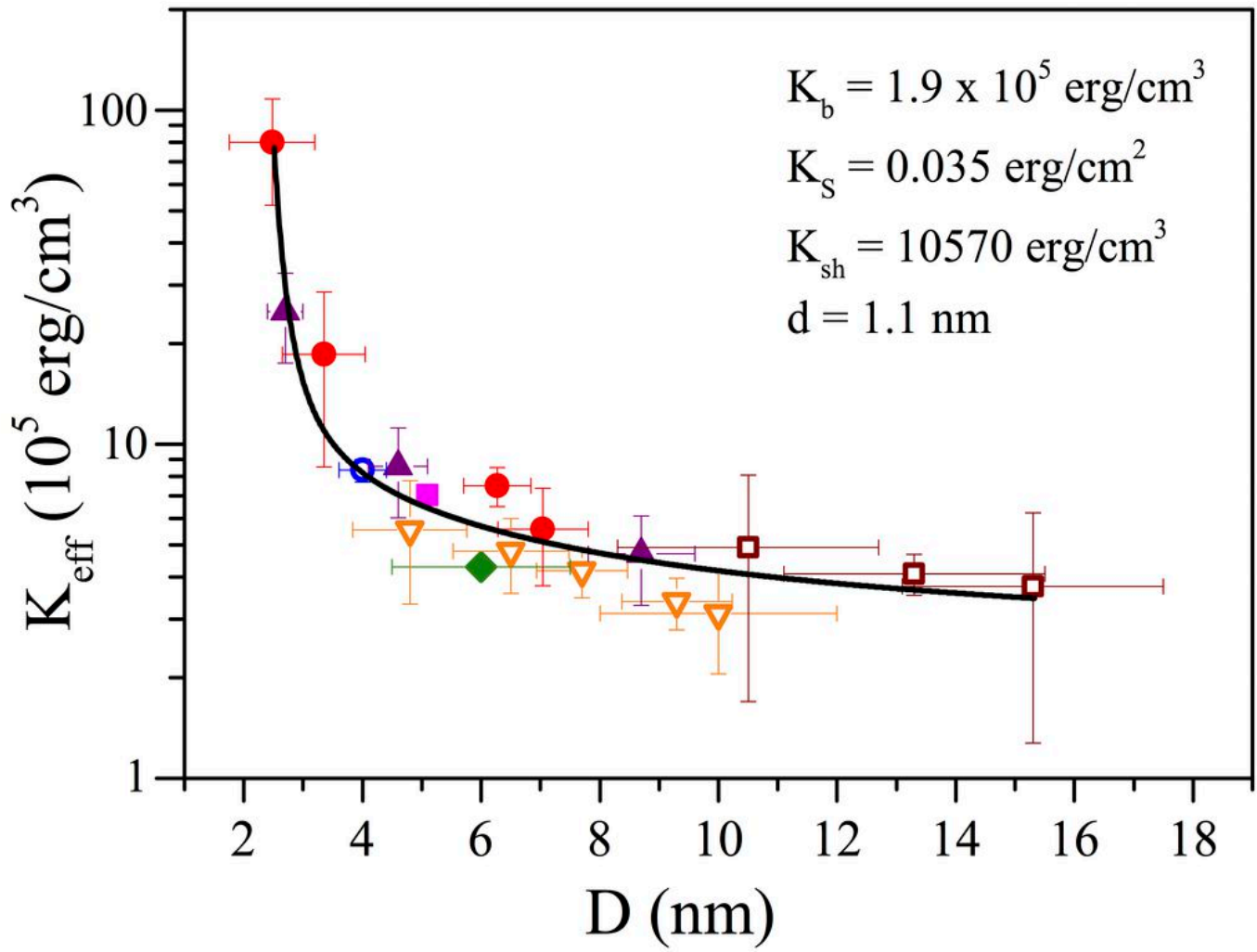

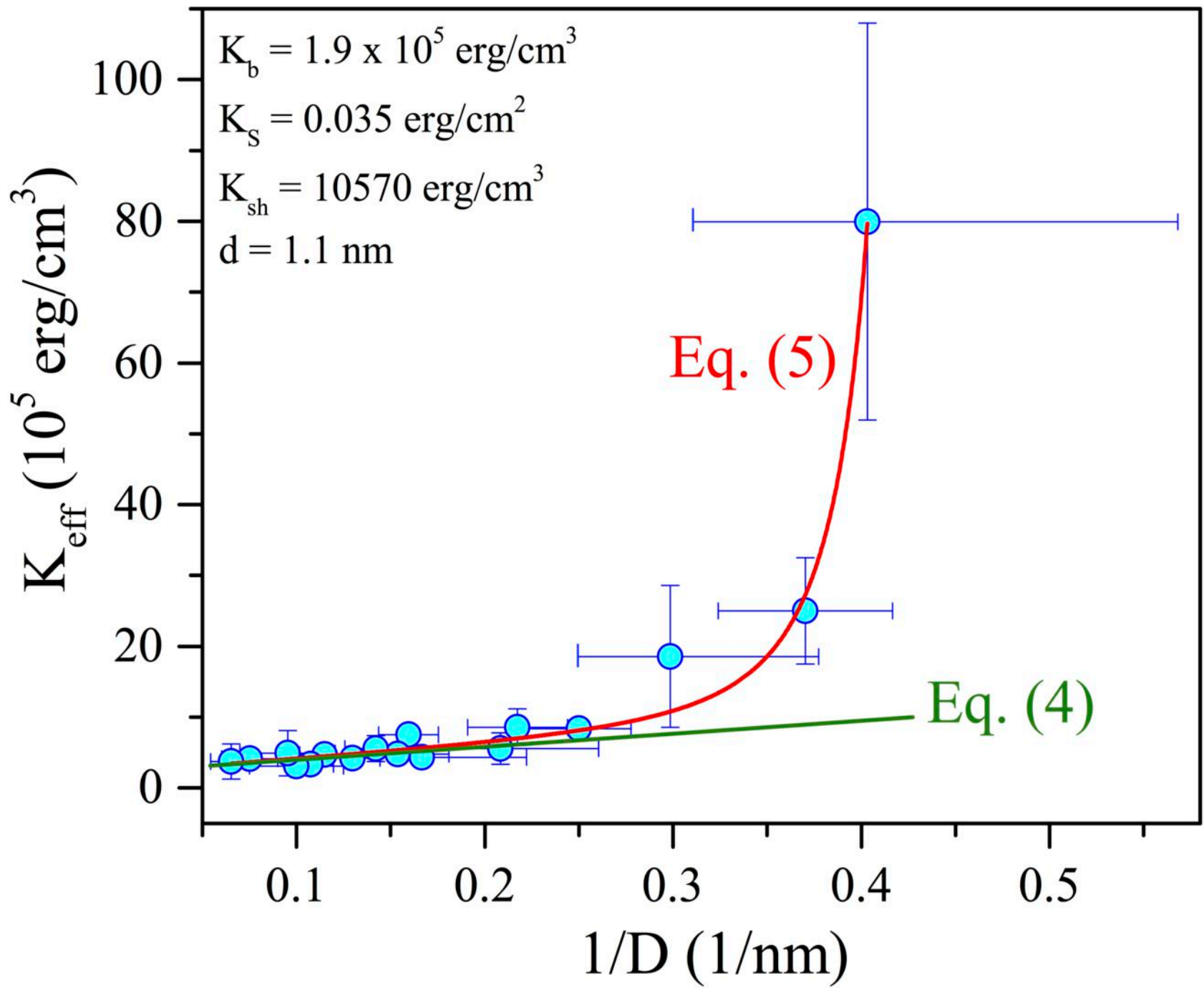

**Supplemental Information: Unusual enhancement of effective magnetic anisotropy with decreasing particle size in maghemite nanoparticles**

K. L. Pisane, Sobhit Singh and M. S. Seehra*

Department of Physics & Astronomy, West Virginia University, Morgantown, WV 26506, USA

---------------------------------------------------------------------------------------------------------

**Experimental Details:**

Nanoparticles of γ-Fe$_2$O$_3$ were synthesized using a modification of a procedure of Hyeon et al. [1] and it is described in detail in [2]. This procedure involves reacting Fe(CO)$_5$ with an appropriate mixture of oleic acid (OA) and dioctyl ether at 100°C to form an iron-oleate complex which is then heated to 285 °C to form iron nano-crystallites. After cooling the mixture to room temperature, dehydrated (CH$_3$)$_3$NO is added before heating to 130 °C for two hours and refluxing at 285 °C for one hour to oxidize the Fe NPs to form γ-Fe$_2$O$_3$. The γ-Fe$_2$O$_3$ NPs are extracted from the solution using ethanol, toluene and a magnet. The size of the particles is controlled by adjusting the weight ratio of Fe(CO)$_5$ to OA. The ratios of Fe(CO)$_5$ to OA used were 2:1, 1:1, 1:3, and 1:4 to produce particles of average size D = 2.5 nm, 3.4 nm, 6.3 nm, and 7.0 nm, respectively.

The synthesized particles were structurally characterized by x-ray diffraction (XRD) using a Rigaku RU-300 x-ray diffractometer with CuKα radiation (λ=0.15418 nm) and transmission electron microscopy (TEM) using a JEOL JEM 2100 system. Fourier transform infrared (FTIR) spectroscopy of the samples was used to ensure that the NP surfaces are coated with oleic acid (OA) [3]. After completing the magnetic measurements, thermogravimetric analysis (TGA) was used to evaporate the OA and determine the mass of the γ-Fe$_2$O$_3$ contained in each sample which is used to determine the magnetization in units of emu/g.

**References:**
1. T. Hyeon, S. S. Le, J. Park, Y. Chung, and H. B. Na, J. Am. Chem. Soc. 123, 12798 (2001)
2. K. L. Pisane, Effects of size and size distribution on the magnetic properties of maghemite nanoparticles and iron-platinum core-shell nanoparticles (Ph.D. dissertation, 2015), West Virginia University, Morgantown, WV, USA

**Supplemental Figures**

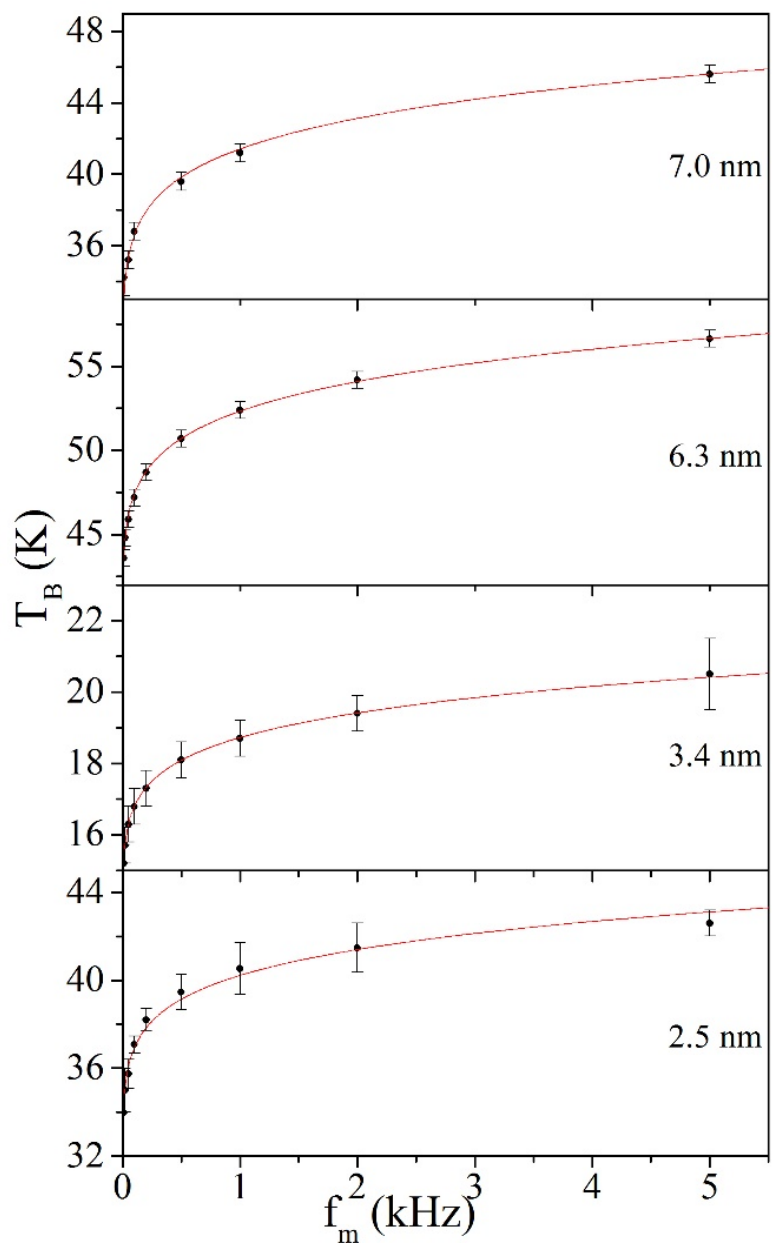

Fig. S1. Frequency dependence of the blocking temperatures $T_B$ for the four $\gamma$-$Fe_2O_3$ NPs samples. The solid lines are fits to the Eq.: $T_B = T_o + (K_{eff}V) / [k_B \ln (f_o/f_m)]$ using $f_o = 2.6 \times 10^{10}$ Hz and $T_o = 0, 11, 2.5$ and $12.5$ K for the D = 7.0, 6.3, 3.4, and 2.5 nm NPs respectively.

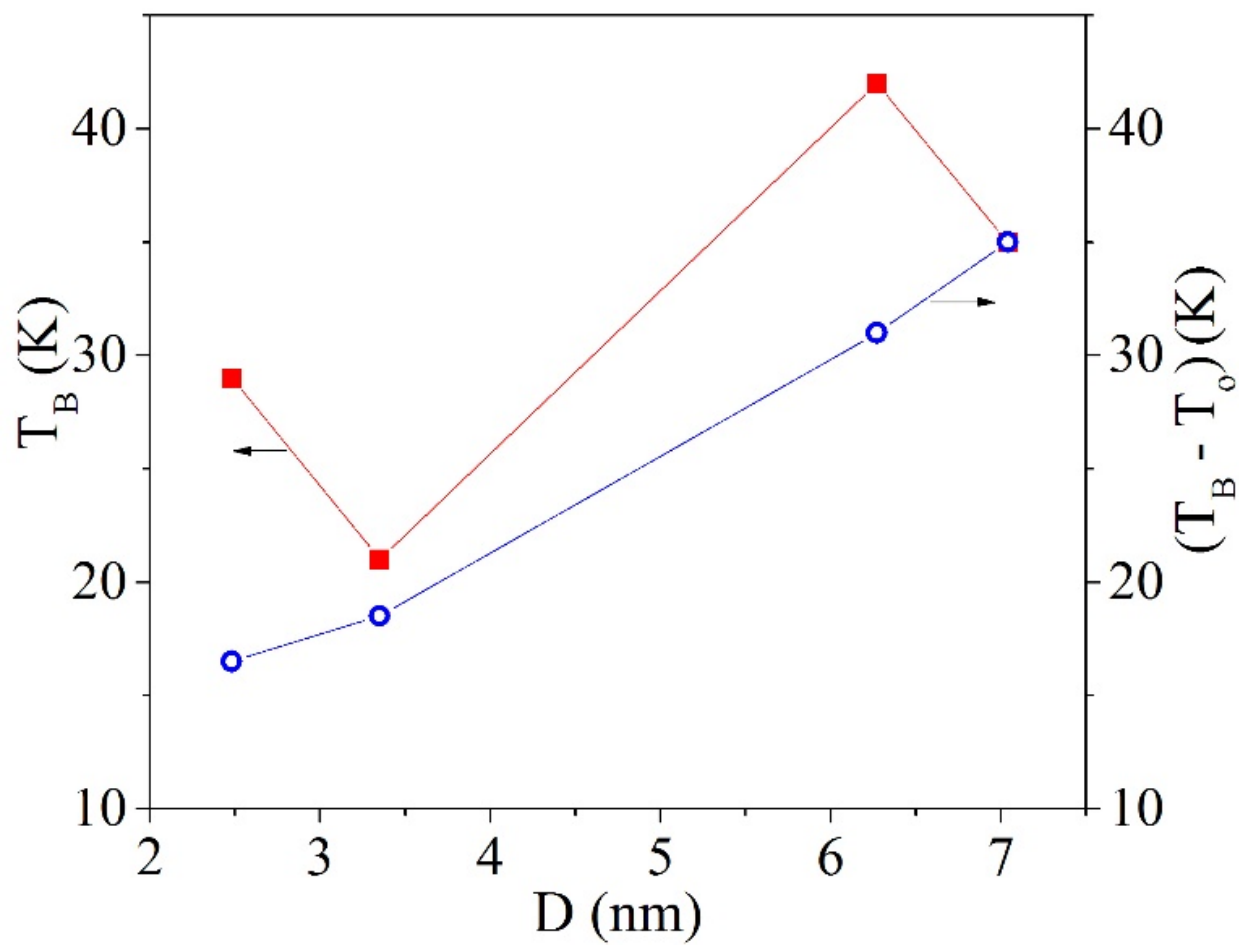

Fig. S2. Plots of the blocking temperature $T_B$ and $(T_B-T_o)$ versus size D of the four γ-Fe$_2$O$_3$ NPs investigated in this work. The lines connecting the data points are visual guides.